\documentclass[aps,epsfig,showpacs,twocolumn]{revtex4-1}
\usepackage{mathrsfs}
\usepackage{amssymb}
\usepackage{graphicx}
\usepackage{epsfig}
\usepackage{amsmath}
\usepackage{amsfonts}
\usepackage{subfigure}

\begin{document}

\title{Probing semiconductor quantum dot state and manipulation with
superconducting transmission line resonator}
\author{Zhe Guan}
\email{guanz@mail.ustc.edu.cn}

\affiliation{Department of Physics,
University of Science and Technology of China, Hefei 230026, People's Republic of
China}

\begin{abstract}
A coupled system of a superconducting transmission line resonator with a
semiconductor double quantum dot is analyzed. We simulate the phase shift of
the microwave signal in the resonator, which is sensitive to the quantum dot
qubit state and manipulation. The measurement quality is sufficiently high
and the results demonstrate a solid-state quantum processor based on this
type of circuit can be envisioned.
\end{abstract}

\date{June 15, 2013}
\maketitle


The interaction between light and matter has been a major focus of research
in atomic physics and quantum optics for a long time \cite{quantum optics}.
There were many significant advances in the realization of quantum phenomena
in manipulating atoms and photons to demonstrate elementary aspects of quantum
physics\cite{wineland86,wineland95,haroche96,haroche92}.
The most fundamental form of this kind of interaction is that a single
photon interacts with a single atom. With the generation of cavity quantum
electrodynamics, coherent quantum Rabi oscillations between the atom and the
cavity has been experimentally observed using Rydberg atoms in microwave
cavities \cite{haroche96,haroche01} and alkali atoms in optical cavities
\cite{kimble92}. Developments have been made in replacing the natural atoms
with artificial two-level systems in solid-state systems, such as the
superconducting quantum bits (qubits) \cite{martinis02,yamamoto03,schoelkopf04}.
 Recently, a semiconductor double quantum dot (DQD) coupled with electromagnetic field of
a superconducting microwave resonator has been experimentally realized \cite%
{ensslin12,komiyama12,petta12}. Especially, the observed distinct vacuum
Rabi splitting indicates that a strong coupling exists in this system \cite%
{komiyama12}. By introducing the architecture of circuit quantum
electrodynamics (cQED), this semiconductor-superconductor hybrid system
shows the advantage of scalability to coherently couple qubits and
resonators on one chip compared to other systems. Furthermore, the coupled system has been
applied to spin dynamics\cite{petta12}. These works have paved a way for fast quantum
information processing in semiconductor structures, which would be rather
useful in quantum computing and quantum communication. Here we demonstrate
how to probe the DQD state and manipulation using the superconducting
transmission line resonator, exploiting the potential of the resonator as
'quantum detector' of the qubit.

\begin{figure}[tbph]
\includegraphics[width=1.0\columnwidth]{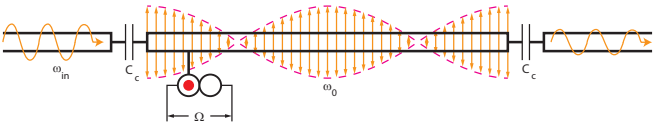}
\caption{The GaAs/AlGaAs semiconductor double quantum dot is coupled to the
superconducting transmission line resonator via capacitive coupling, and a
continuous microwave is induced to the transmission line from the left side.}
\end{figure}

The schematic diagram of the capacitive coupled system of GaAs/AlGaAs
semiconductor DQD and superconducting transmission line resonator, is shown
in Fig. 1. Jaynes-Cummings Hamiltonian can be used to describe the whole
system under the rotating-wave approximation \cite{ensslin12}:

\begin{equation}
\begin{split}
& \hat{H}_0=\frac{\hbar \Omega }{2}\hat{\sigma}_{z}+\hbar \omega _{0}(\hat{a}%
^{\dagger }\hat{a}+\frac{1}{2})+\hbar g(\hat{\sigma}_{+}\hat{a}+\hat{\sigma}%
_{-}\hat{a}^{\dagger }), \\
& \Omega =\sqrt{\varepsilon ^{2}+4t^{2}}
\end{split}%
\end{equation}%
Here $\hat{\sigma} _{z}$, $\hat{\sigma} _{+}$, $\hat{\sigma} _{-}$ are the Pauli operators in
the basis of $|\uparrow \rangle $ (qubit excited state) and $|\downarrow
\rangle $ (qubit ground state) of the DQD, $a$ and $a^{\dagger }$ are
the annihilation and creation operators of the photon in the resonator
respectively, and  $\omega _{0}$ is the bare frequency of the resonator.
$\varepsilon $ is the detuning energy between the charge qubit
states, $t$ is the inter-dot tunneling rate, and $g$ is the coupling
strength between the DQD and the resonator.

Currently research on this system focuses on the dispersive region \cite
{ensslin12,petta12}. To realize probing the DQD state using the superconducting transmission
line, we can simulate the time-evolution of the density matrix of the
coupled system, with the transmission line imposed by a continuous microwave
field. We use the standard master-equation method with a Markovian
approximation \cite{ensslin12}:

\begin{equation}
\begin{split}
\frac{d\rho }{dt}=& \frac{i}{\hbar }[\rho ,\hat{H}_{0}]-\frac{\kappa }{2}(%
\hat{a}^{\dagger }\hat{a}\rho -2\hat{a}\rho \hat{a}^{\dagger }+\rho \hat{a}%
^{\dagger }\hat{a}) \\
& -\frac{\gamma _{l}}{2}(\hat{\sigma}_{-}\hat{\sigma}_{+}\rho -2\hat{\sigma}%
_{-}\rho \hat{\sigma}_{+}+\rho \hat{\sigma}_{-}\hat{\sigma}_{+}) \\
&-\frac{%
\gamma _{\phi }}{2}(\rho -\hat{\sigma}_{z}\rho \hat{\sigma}_{z}+\rho)
\end{split}%
\end{equation}%
Here $\rho $ is the density matrix of the coupled system, $\kappa $ is the
decay rate of the resonator, $\gamma _{l}$ is the energy relaxation rate, $%
\gamma _{\phi }$ is the dephasing rate, $\omega _{0}$ is the frequency of
the resonator. When the DQD is near the inter-dot charge transitions ($%
\varepsilon =0$), the energy of the DQD is close the microwave field energy,
allowing a more effective photon exchange between the DQD and the microwave
field. The strengthened exchange results in a phase shift of the microwave
imposed in the transmission line. To calculate the phase shift under various
circumstances of the coupled system, we use $\phi =\arg (i\left<\hat{a}\right>)$
as the phase response \cite{Wallraff05, Schoelkopf04a}.  In the Shr\"{o}dinger picture, we could
 easily extract the phase shift of the microwave after we simulate the time-evolution
 of the density matrix of the coupled system. Especially, we have the phase shift $
\triangle \phi =-\mathrm{{arctan}(2g^{2}/\kappa \Delta )}$ when $\Delta >g$.

In Fig. 2, the numerical simulation of phase shift $\triangle \phi $ of the
microwave is plotted as a function of $\varepsilon $ and $2t$ of the DQD,
with $2t/\hbar $ ranging from $0$ to $20$ GHz and $\varepsilon /\hbar $
ranging from $-20$ to $20$ GHz. Here we use the realistic parameters in the
experiments as $\omega _{0}/2\pi=6.2$ GHz, $\kappa/2\pi =3.1$ MHz, $\gamma _{l}=66.7$ MHz, $\gamma _{\phi }=0$ MHz
\cite{ensslin12,komiyama12,petta12}.  Sign change of the DQD-transmission line detuning, $\Delta$ at different
value of $\varepsilon$ introduces the sign change of $\triangle\phi$. Along the $2t$ axis at a fixed $%
\varepsilon $, there is a sharp increment of $\triangle \phi $ from negative
to positive near $2t/\hbar =$ $6.2$ GHz, which could be clearly observed in
the 3D plot of phase shift distribution in the $\varepsilon -2t$ plane. From
this picture, we can conclude that the DQD spectroscopy can be efficiently
probed by the coupled transmission line resonator.

\begin{figure}[tbph]
\includegraphics[width=1.0\columnwidth]{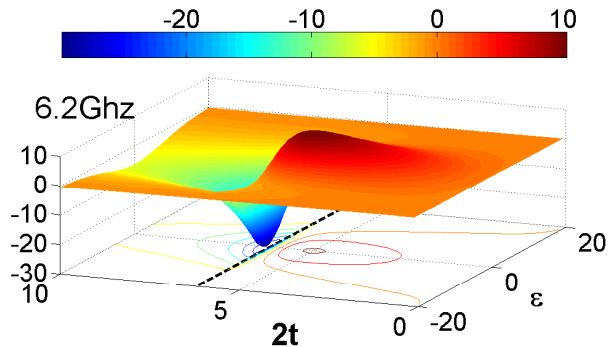}
\caption{The phase shift $\triangle \protect\phi $ of the microwave is
depicted as the function of $\protect\varepsilon $ and $2t$. In the $\varepsilon-2t$ plane, the density of isolines when $2t/\hbar$ is near 6.2 GHz is much bigger than other part of the plane, illustrating the sharp increment of $\triangle\phi$. }
\end{figure}

\begin{figure}[tbph]
\subfigure[]{\label{3a}\includegraphics[width=1.0\columnwidth]{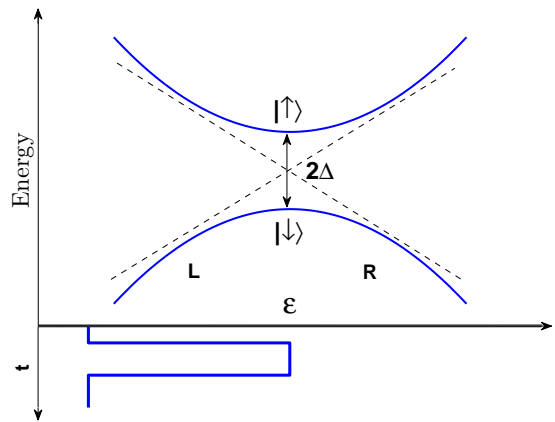}}
\subfigure[]{\label{3b}\includegraphics[width=1.0\columnwidth]{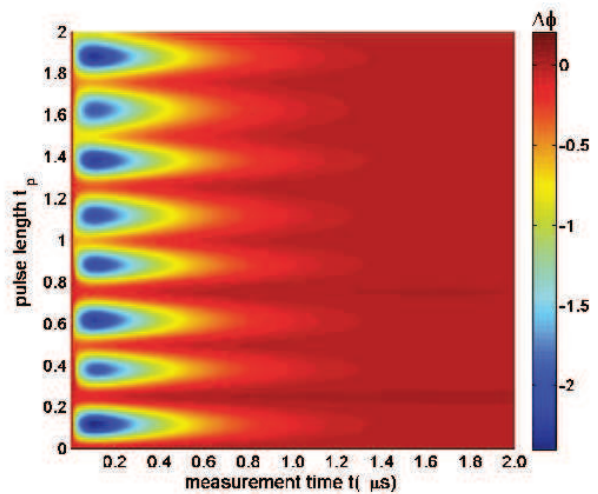}}
\subfigure[]{\label{3b}\includegraphics[width=1.0\columnwidth]{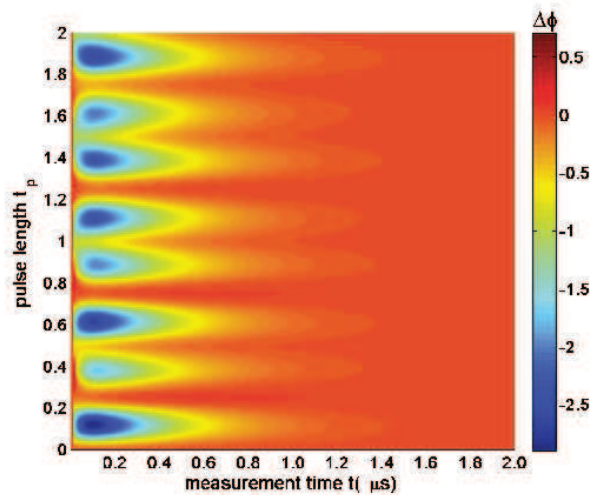}}
\caption{ (a) Schematic of the charge qubit energy level diagram with a short rectangle driven pulse.
 (b) The phase shift $\triangle \protect\phi $ of the microwave
resonator is depicted as a function of measurement time $t$ and pulse length
$t_{p}$. The photon number in the resonator is 3.8. A clear qubit Rabi oscillation pattern with little deviation is obtained in the phase shift domain. (c)When the photon in the resonator is 0.6, the less frequency modes would increase the disturbing effect on the original manipulation frequecy, which induces a lager deviation from perfect Rabi oscillation pattern.}
\end{figure}

We can also probe the manipulation process of the semiconductor DQD by
reading out the phase shift of the microwave. In this scheme, an operation
pulse is applied to the DQD as illustrated in Fig. 3a \cite{petta10}. The
DQD initially stays at its charge state $\left\vert R\right\rangle $. Due to
the employed short pulse, the DQD will be driven to the charge degeneracy
point nonadiabatically. Since $\left\vert R\right\rangle $ is no longer an
eigenstate, the DQD starts to evolve during the pulse duration time $t_{p}$.
When the pulse is terminated, the DQD stays in the superposition of $%
\left\vert R\right\rangle $ and $\left\vert L\right\rangle $ states. After
the manipulation pulse is applied to the DQD, the energy relaxation of the
qubit exists and we conduct continuous monition of the resulted phase shift
of the resonator.

In Fig. 3b and 3c, we plot the phase shift $\triangle \phi $ as a function of
measurement time $t$ and pulse length $t_{p}$ with different amplitudes of the
induced microwave. Here we use the realistic parameters as $\omega _{0}/2\pi=6.2$ GHz,
$\kappa/2\pi =1$ MHz, $\gamma _{l}=20$ MHz, $\gamma _{\phi }=200$ MHz \cite{ensslin12,komiyama12,petta12}.
In Fig. 3b, the steady photon number in the resonator is $3.8$, we can see that
when apply a short rectangle pulse to manipulate the DQD, we can obtain a clear
periodical Rabi oscillation pattern of the phase shift. Before the pulse is applied,
the phase shift has a minimum value which is approximately $0$ corresponding to
the initial state of the qubit. When $t_{p}=0.1$ ns, we have the maximum phase
 shift of $-2.4 $ deg, and this is because the qubit has been pulled to its superposed
state due to the induced pulse. On the contrary, when $t_{p}=0.3$ ns, no phase
shift is observed since the qubit remains its initial state under the pulse.
Also the phase shift response to the $t_{p}=0.1$ ns pulse is almost the same
with that to the $t_{p}=0.6$ ns pulse, and this is due to that the time scale
of Rabi oscillation is much shorter than the energy relaxation time of the
qubit. Additionally, since the qubit would decay from its excited state to
the ground state exponentially with the relaxation time $T_{1}=1/\gamma $,
the phase shift would also decay as the measurement time $t$ extends, and it
is observable that after $t=0.7$ $\mu $s, hardly we can monitor any phase
shift even though a pulse is applied on the qubit. Also, we could obviously sense the
phenomenon that the phase shift Rabi oscillation pattern is not perfectly periodic.
The reason lies in that our manipulation pulse length is small thus the disturbance from
other frequency modes besides the manipulation frequency is comparably notable, which
results in the deviation from periodicity. The deviation is more serious in Fig. 3c where
the steady photon number in the resonator is 0.6, which means the frequency modes in the system
is less than that in Fig. 3b. The mix of more modes would decrease the disturbing effect, so it
is not hard for us to understand the periodicity is worse in Fig. 3c. Overall, we have achieved
probing the DQD manipulation by monitoring the phase shift of the microwave
resonator.

In conclusion, we have comprehensively analyzed and simulated the phase
shift in microwave transmission line, which could help us probe the state of the
DQD. Additionally, by monitoring the phase shift of the microwave transmitted in
the line, we could probe our operation on the DQD.
Our work paves a way to possible applications
using the resonator as the highly efficient detector which would be
attractive to the further quantum information processing.

This work was supported by the National Basic Research Program of China (Grant
No.~ 2011CB921200), the CAS, the National Natural Science Foundation of China
(Grant Nos. 11274289 and 11105135), the Fundamental Research Funds for the
Central Universities (Nos. WK2470000011). ZG acknowledges support from the Fund
for Fostering Talents in Basic Science of the National Natural Science Foundation
of China (No.~J1103207).

\end{document}